Refractive index sensing with temperature compensation by a multimode-interference fiber-based optical frequency comb sensing cavity

Ryo Oe,[1,2] Takeo Minamikawa,[2,3,4] Shuji Taue,[5] Hidenori Koresawa,[1,2] Takahiko Mizuno,[2,3,4] Masatomo Yamagiwa,[2,3,4] Yasuhiro Mizutani,[2,6] Hirotsugu Yamamoto,[2,7] Tetsuo Iwata,[2,4] and Takeshi Yasui[2,3,4*]

[1]Graduate School of Advanced Technology and Science, Tokushima University, 2-1, Minami-Josanjima, Tokushima, Tokushima 770-8506, Japan

[2]JST, ERATO, MINOSHIMA Intelligent Optical Synthesizer Project, 2-1, Minami-Josanjima, Tokushima, Tokushima 770-8506, Japan

[3]Institute of Post-LED Photonics (pLED), Tokushima University, 2-1, Minami-Josanjima, Tokushima, Tokushima 770-8506, Japan

[4]Graduate School of Technology, Industrial and Social Sciences, Tokushima University, 2-1, Minami-Josanjima, Tokushima, Tokushima 770-8506, Japan

[5]Electronic and Photonic Systems Engineering Course, Kochi University of Technology, 2-22 Eikokuji, Kochi, Kochi 780-8515, Japan

[6]Graduate School of Engineering, Osaka University, 2-1, Yamadaoka, Suita, Osaka 565-0871, Japan

[7]Center for Optical Research and Education, Utsunomiya University, 7-1-2, Yoto, Utsunomiya, Tochigi 321-8585, Japan

**Abstract**

We proposed a refractive index (RI) sensing method with temperature compensation by using an optical frequency comb (OFC) sensing cavity employing a multimode-interference (MMI) fiber, namely, the MMI-OFC sensing cavity. The MMI-OFC sensing cavity enables simultaneous measurement of material-dependent RI and sample temperature by decoding from the comb spacing frequency shift and the wavelength shift of the OFC. We realized the simultaneous and continuous measurement of RI-related concentration of a liquid sample and its temperature with precisions of $1.6 \times 10^{-4}$ RIU and 0.08 ºC. The proposed method would be a useful means for the various applications based on RI sensing.

# 1. Introduction

Refractive index (RI) is an important physical quantity of materials. Furthermore, RI is a useful parameter for the characterization and identification of materials, such as for the determination of the concentrations of liquid solutions [1], molecular identification [2] and biosensing [3]. The fiber-based RI sensor is a powerful means for RI measurement because of its compactness, simplicity, flexibility, noise robustness and availability in various environments and has been applied in many applications in sucrose sensing [4], gas pressure sensing [5] and biomolecule sensing [6]. In the conventional RI fiber sensors, which are based, for example, on surface plasmon resonance (SPR) [7], interference with a core-offset fiber [8] or multimode interference (MMI) [9], the change of RI is measured by a shift of transmitted optical spectrum. However, the sensitivity of the conventional RI sensing is limited due to the limited spectral bandwidth of a light source and spectral resolution of a spectrometer.

To overcome this issue, RI fiber sensors based on radio-frequency (RF) signal measurement have been developed, in which the RI is retrieved by an RI-dependent optical beat signal generated using, for example, fiber Bragg gratings (FBGs) [10], Fabry-Perot interferometers [11], and chip-scale microring resonators [12]. To obtain a high-quality optical beat signal for highly sensitive RI measurement, we recently proposed an optical-frequency comb (OFC)-based RI sensor [13]. The OFC [14-16] is regarded as a group of a vast number of phase-locked narrow-linewidth continuous-wave (CW) light sources with a constant frequency spacing of $f_{rep}$ (typically, 50–100 MHz) over a broad spectral range. The OFC is widely applied for high precision spectroscopy or measurement; for instance, atomic spectroscopy [16], gas spectroscopy [17], solid spectroscopy [18], spectroscopic ellipsometry [19], strain sensing [20,21], ultrasonic-wave sensing [22], distance measurement [23, 24], optical microscopy [25-28], and 3D shape measurement [29].

On the other hand, the OFC has an interesting feature: a photonic RF conversion function in a fiber OFC cavity. Our previous study [13] revealed that the OFC-based RI sensing can achieve an RI resolution of $4.88 \times 10^{-6}$ RIU and RI accuracy of $5.35 \times 10^{-5}$ RIU by the combination with an MMI fiber sensor incorporated in the OFC cavity, namely, an MMI-OFC sensing cavity, which has accomplished an equivalent or higher sensitivity compared to the conventional RI-sensing fiber sensor.

Although the highly sensitive RI measurement was realized based on the MMI-OFC sensing cavity, further improvements of the sensitivity and stability of RI measurement was expected if the sample temperature fluctuation was stabilized or monitored. This is because the RI of a sample is dependent on the temperature of the sample, and the instability of the fiber structure at the sensing region due to temperature fluctuations led to the instability of RI measurement with the MMI-OFC sensing cavity.

In this study, to improve the RI measurement with the MMI-OFC sensing cavity, we realized simultaneous measurement of temperature with the same configuration of MMI-OFC-sensing cavity for RI measurement by simultaneous monitoring of an optical spectral shift in the optical frequency region and a comb spacing shift in the RF region of the OFC. We confirmed the fundamental sensitivity of the MMI-OFC sensing cavity depending on sample temperature and RI-related concentration of a liquid sample. Furthermore, we demonstrated simultaneous and continuous measurement of temperature and RI-related concentration of a liquid sample.

## 2. Principle operation

Figure 1 shows the configuration of an MMI-OFC sensing cavity. The MMI-OFC sensing cavity is comprised of a fiber OFC ring cavity with an intracavity MMI fiber sensor. The intracavity MMI fiber sensor acts as a RI-dependent tunable optical bandpass filter (bandpass center wavelength = $\lambda_{MMI}$). Thus, the optical spectrum of the OFC shows the RI-dependent shift owing to the intracavity MMI fiber sensor, namely, $\lambda_{MMI}$ shift ($\Delta\lambda_{MMI}$).

On the other hand, the OFC exhibited the comb-tooth-like ultradiscrete multimode spectrum with a constant frequency spacing $f_{rep}$ given by,

$$f_{rep} = \frac{c}{nL}, \quad (1)$$

where $c$ is the speed of light in vacuum, $n$ is the group RI of the cavity fiber, and $L$ is the geometrical length of the fiber cavity. Since $f_{rep}$ is determined by the optical cavity length $nL$, it is sensitive to the cavity disturbance changing $nL$; in other words, $f_{rep}$ can be used for quantitative analysis of a physical quantity acting as the cavity disturbance. The intracavity MMI fiber sensor will change the optical cavity length at the intracavity MMI fiber sensor and the group RI of the cavity fiber owing to the spectral shift of the OFC, leading to the RI-dependent $f_{rep}$ shift, namely, $\Delta f_{rep}$.

$\lambda_{MMI}$ and $f_{\text{rep}}$ are shifted by a temperature change of a sample because the temporal change leads to a change of sample RI and/or changes of RI and geometrical shape in the MMI sensor fiber. Thus, by observing $\Delta\lambda_{MMI}$ and $\Delta f_{rep}$, we consider realizing the simultaneous measurement of the RI and sample temperature as follows,

$$\begin{bmatrix} \Delta f_{rep} \\ \Delta\lambda_{MMI} \end{bmatrix} = \begin{bmatrix} \frac{\partial f_{rep}}{\partial C} & \frac{\partial f_{rep}}{\partial T} \\ \frac{\partial \lambda_{MMI}}{\partial C} & \frac{\partial \lambda_{MMI}}{\partial T} \end{bmatrix} \begin{bmatrix} \Delta C \\ \Delta T \end{bmatrix}, \quad (2)$$

where $\Delta C$ and $\Delta T$ are RI-related sample parameters, *i.e.*, concentration and sample temperature, respectively. The coefficient matrix is comprised of the slope coefficients between $\Delta\lambda_{MMI}$ or $\Delta f_{rep}$ and $\Delta C$ or $\Delta T$. From Eq. 2, if the coefficient matrix is regular, one can determine $\Delta C$ and $\Delta T$ from the simultaneous measurement of $\Delta\lambda_{MMI}$ and $\Delta f_{rep}$ and the inverse matrix calculation of them.

To demonstrate the MMI-OFC sensing cavity with the temperature compensated RI sensing, a custom-built ring-cavity erbium-doped fiber (EDF) laser with an intracavity MMI fiber sensor, which was mode-locked by a nonlinear polarization rotation technique, was employed as an OFC generator (the fundamental comb spacing or repetition rate, 43 MHz; the center wavelength, 1555 nm; and the spectral bandwidth, 15 nm). The intracavity MMI-fiber sensor was made with a core-less multimode fiber (MMF, the core diameter, 125 μm; the length, 58 mm, FG125LA, Thorlabs) spliced with single-mode fibers (SMF-28-100, Thorlabs) at each terminal. The intracavity MMI-fiber sensor was sunk in a liquid sample with a glass tube. The glass tube had an inlet port and an outlet port to externally change and stir the liquid sample. $\Delta\lambda_{MMI}$ or $\Delta f_{rep}$ were obtained with an optical spectrum analyzer (OSA, AQ6315A, Yokogawa Electric Corp.), an RF spectrum analyzer (RFSA, E4402B, Keysight Technologies), or an RF frequency counter (53230A, Keysight Technologies).

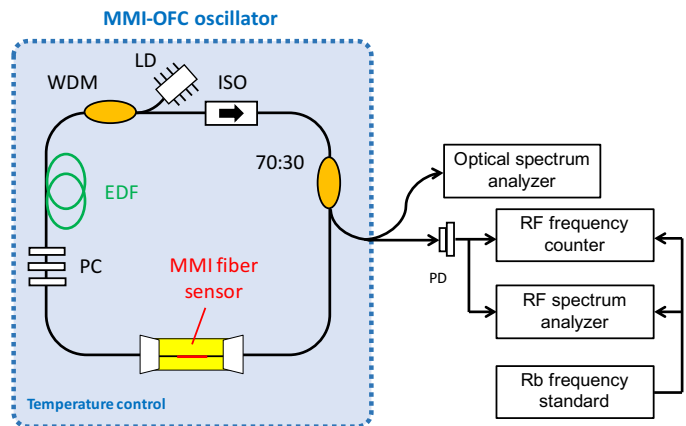

Fig. 1. Schematic diagram of the MMI-OFC. PC, polarization controller for mode-locking; EDF, Er-doped fiber; WDM, wavelength division multiplexer; ISO, polarization independent isolator; 70:30, branch coupler; PC, polarization controller; LD, laser diode; PD, photo detector.

## 3. Results

### *3.1 Temperature dependencies of $\Delta\lambda_{MMI}$ and $\Delta f_{rep}$ in the MMI-OFC sensing cavity*

We first investigated $\Delta\lambda_{MMI}$ and $\Delta f_{rep}$ with respect to different sample temperatures. The ethanol/water solution (ethanol concentration of 0−15% v/v with interval of 5% v/v) was used for samples with different temperatures (26.5−29.0ºC with interval of 0.5ºC). The same experiment was repeated for 3 sets of ethanol/water samples with each temperature, and their standard deviations under each of the samples' temperatures were determined.

Figure 2a shows the temperature-dependent optical spectra of the MMI-OFC sensing cavity for the ethanol concentration of 0% v/v. Figure 2b shows the relation between sample temperature and $\Delta\lambda_{MMI}$ with respect to different ethanol concentrations. We confirmed the temperature-dependent increase of $\Delta\lambda_{MMI}$. The slope coefficients between temperature and $\Delta\lambda_{MMI}$ were determined to be 0.016 nm/ºC for 0% v/v, 0.022 nm/ºC for 5% v/v, 0.021 nm/ºC for 10% v/v, and 0.018 nm/ºC for 15% v/v. Thus, the mean slope coefficient, namely, $\partial\lambda_{MMI}/\partial T$, was obtained to be 0.020 nm/ºC.

Figure 2c shows the temperature-dependent RF spectra of the MMI-OFC sensing cavity for the ethanol concentration of 0% v/v. Figure 2d shows the relation between sample temperature and $\Delta f_{rep}$ with respect to different ethanol concentrations. We confirmed the temperature-dependent decrease of $\Delta f_{rep}$. The slope coefficients between temperature and $\Delta f_{rep}$ were determined to be -18.01 Hz/ºC for 0% v/v, -23.63 Hz/ºC for 5% v/v, -19.74 Hz/ºC for 10% v/v, and -19.00 Hz/ºC for 15% v/v%, respectively. Thus, the mean slope coefficient, namely, $\partial f_{rep}/\partial T$, was found to be -20.46 Hz/ºC. An opposite sign of slope coefficient between the temperature-dependent $\Delta\lambda_{MMI}$ and $\Delta f_{rep}$ was obtained.

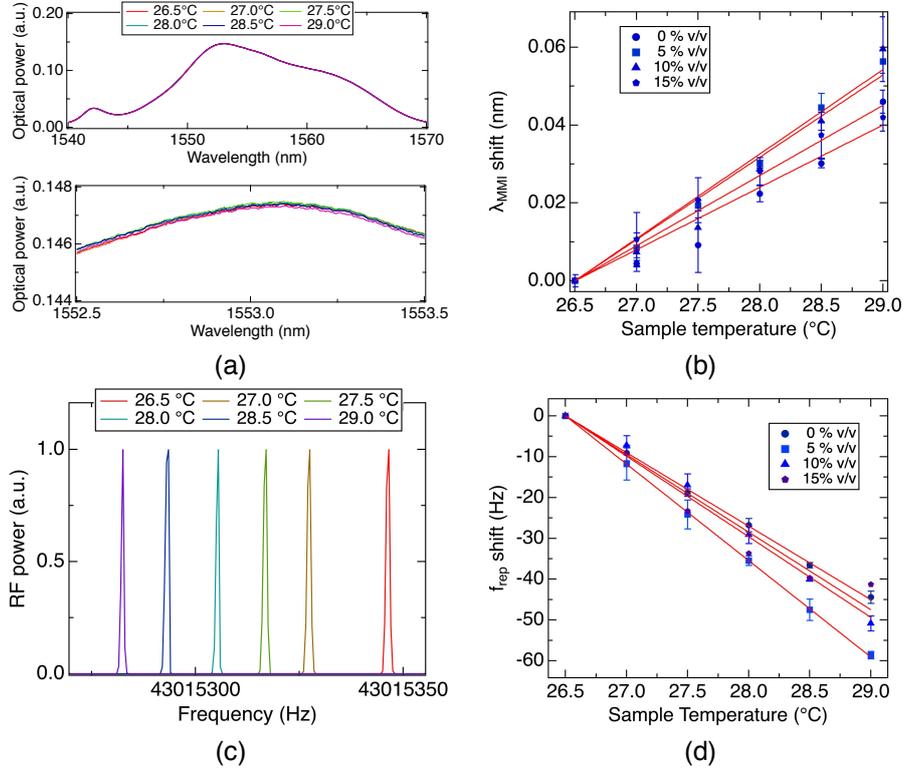

Fig. 2. Temperature-dependent optical spectral shift and RF spectral shift. (a) Optical spectral shift with respect to temperature change. (b) Temperature sensitivity of optical spectrum under constant sample concentration. (c) RF spectral shift with respect to temperature change. (d) Temperature sensitivity of RF spectrum under constant sample concentration

### *3.2 RI-related concentration dependencies of Δλ$_{MMI}$ and Δf$_{rep}$ in the MMI-OFC sensing cavity*

We next investigated Δλ$_{MMI}$ and Δf$_{rep}$ with respect to material-dependent RI for which the different ethanol concentrations of the liquid sample were used. The ethanol/water solution (ethanol concentrations of 0−15% v/v with interval of 3% v/v) was used for samples with different temperatures (26.0−29.0ºC with interval of 1ºC). The same experiment was repeated for 3 sets of ethanol/water samples with each concentration, and their standard deviations were determined for each of the sample concentrations.

Figure 3a shows the concentration-dependent optical spectra of the MMI-OFC sensing cavity at a constant temperature of 26ºC. Figure 3b shows the relation between sample concentration and $\Delta\lambda_{MMI}$ with respect to different sample temperatures.  We confirmed the concentration-dependent increase of $\Delta\lambda_{MMI}$. The slope coefficients between temperature and $\Delta\lambda_{MMI}$ were determined to be 0.029 nm/% v/v for 26.0ºC, 0.034 nm/% v/v for 27.0ºC, 0.028 nm/% v/v for 28.0ºC, and 0.035 nm/% v/v for 29.0ºC. Thus, the mean slope coefficient, namely, $\partial\lambda_{MMI}/\partial C$, was obtained to be 0.0315 nm/% v/v.

Figure 3c shows the concentration-dependent RF spectra of the MMI-OFC sensing cavity at a constant temperature of 26ºC. Figure 3d shows the relation between sample concentration and $\Delta f_{rep}$ with respect to different sample temperatures. We confirmed the temperature-dependent decrease of $\Delta f_{rep}$. The slope coefficients between temperature and $\Delta f_{rep}$ were determined to be -3.10 Hz/% v/v for 26.0ºC, -2.66 Hz/% v/v for 27.0ºC, -3.05 Hz/% v/v for 28.0ºC, and -3.03 Hz/% v/v for 29.0ºC. Thus, the mean slope coefficient, namely, $\partial f_{rep}/\partial C$, was obtained to be -2.96 Hz/% v/v. An opposite sign of the slope coefficient between the temperature-dependent $\Delta\lambda_{MMI}$ and $\Delta f_{rep}$ was confirmed again.

According to the results of the temperature and RI-related concentration dependencies, we can estimate the precision of the MMI-OFC sensing cavity. The obtained mean stabilities of $\Delta\lambda_{MMI}$ and $\Delta f_{rep}$ defined by the standard deviation were $8.9\times10^{-2}$ nm and 0.70 Hz, respectively. As a result, the precisions of the temperature and RI-related concentration measurements with the MMI-OFC sensing cavity were 0.08 ºC and 0.34 % v/v corresponding to $1.6\times10^{-4}$ RIU, respectively. This result indicates that the MMI-OFC sensing cavity has the potential for RI sensing with temperature compensation with high precision.

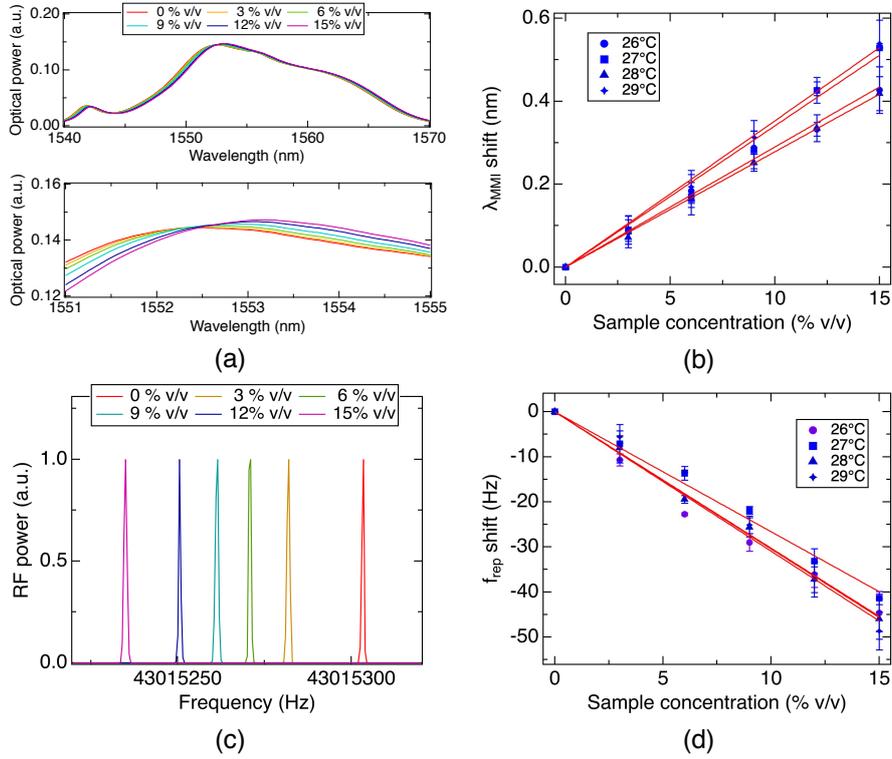

Fig. 3. Concentration-dependent optical spectral shift and RF spectral shift. (a) Optical spectral shift with respect to sample ethanol-concentration change. (b) Ethanol-concentration sensitivity of optical spectrum under constant sample temperature. (c) RF spectral shift with respect to ethanol-concentration change. (d) Ethanol-concentration sensitivity of RF spectrum under constant sample temperature.

*3.3 Simultaneous measurement of temperature and concentration of a liquid sample*

To highlight the proposed method, we demonstrated simultaneous measurement of temperature and RI-related concentration of an ethanol/water solution. We first increase the temperature of a sample stepwise from 26.5ºC to 29.0ºC by a heater attached to the liquid sample tube under a constant ethanol concentration of 0% v/v. We also gently stirred the sample by flowing it in and out using a syringe after changing the temperature to achieve a uniform temperature of the sample. As shown in Figs. 4a and 4b, we clearly confirmed the temperature-dependent $\Delta\lambda_{MMI}$ increase and $\Delta f_{rep}$ decrease. Then, we determined $\Delta C$ and $\Delta T$ by substituting the measured $\Delta\lambda_{MMI}$ and $\Delta f_{rep}$ for Eq. 3 as shown in Figs. 4c and 4d. The determined $\Delta C$ and $\Delta T$ values well reflect their actual changes: stable concentration within 1% v/v and step-by-step increase of temperature by 0.5ºC within the deviation of below 0.3ºC from the actual change. If the temperature compensation was not performed, in which the $\Delta n$ was determined with only $\Delta f_{rep}$, the estimation error of $\Delta C$ was degraded to 16.9% v/v by the temperature change from 26.5ºC to 29.0ºC, as shown by the green line in Fig. 4.

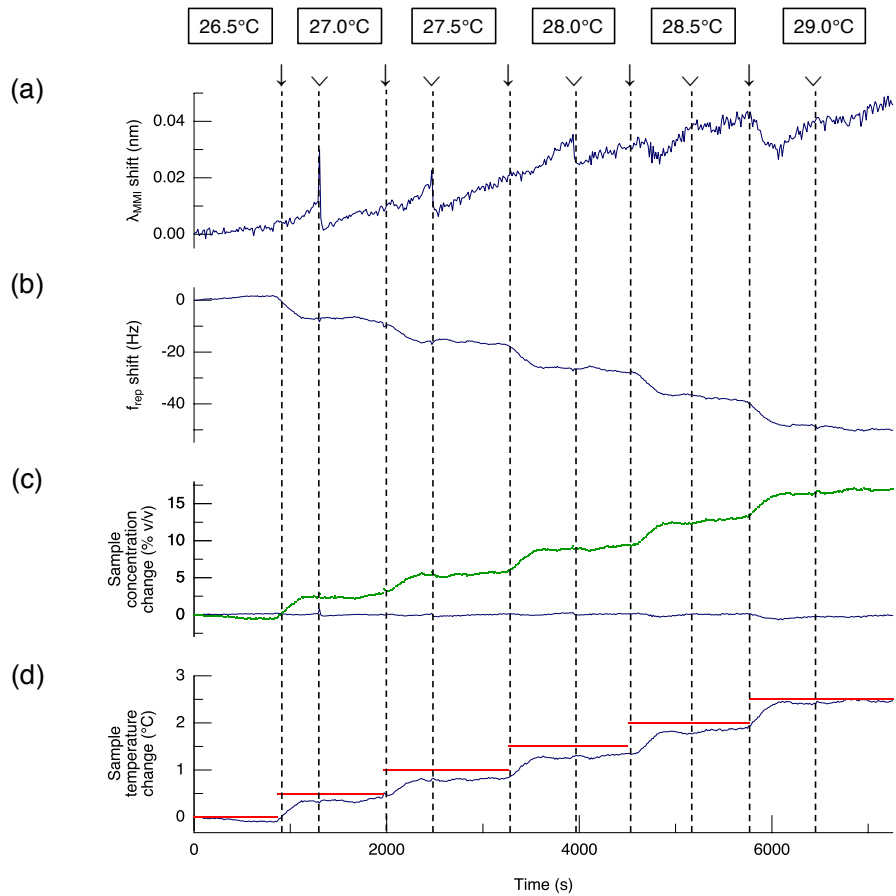

Fig. 4. Temporal behaviors of the $\lambda_{MMI}$ shift, $f_{rep}$ shift, the estimated sample concentration change and the estimated temperature change under the variable temperature (26.5-29.0ºC) and the constant ethanol concentration (0% v/v). The sample concentration change indicated by the green line was estimated by only using $\Delta f_{rep}$ for comparison. The arrows indicate the times at which the temperature was changed. The arrow heads indicate the times at which the sample was stirred.

We next increase the sample concentration stepwise from 0% v/v to 15% v/v by adding ethanol with a pipette under a constant temperature (26.0ºC). We also gently stirred the sample by flowing it in and out using a syringe after the ethanol application to achieve a uniform sample concentration. As shown in Figs. 5a and 5b, we clearly confirmed the concentration-dependent $\Delta\lambda_{MMI}$ increase and $\Delta f_{rep}$ decrease. Then, we determined $\Delta C$ and $\Delta T$ by substituting the measured $\Delta\lambda_{MMI}$ and $\Delta f_{rep}$ for Eq. 3 as shown in Figs. 5c and 5d. The determined $\Delta C$ and $\Delta T$ values again well reflect their actual changes. The measurement stability was approximately 1% v/v in the concentration estimation and below 0.2ºC in the temperature estimation. Importantly, even in the fixed temperature condition, the temperature instability due to the limited control capability of the sample temperature can be compensated by using the proposed method.

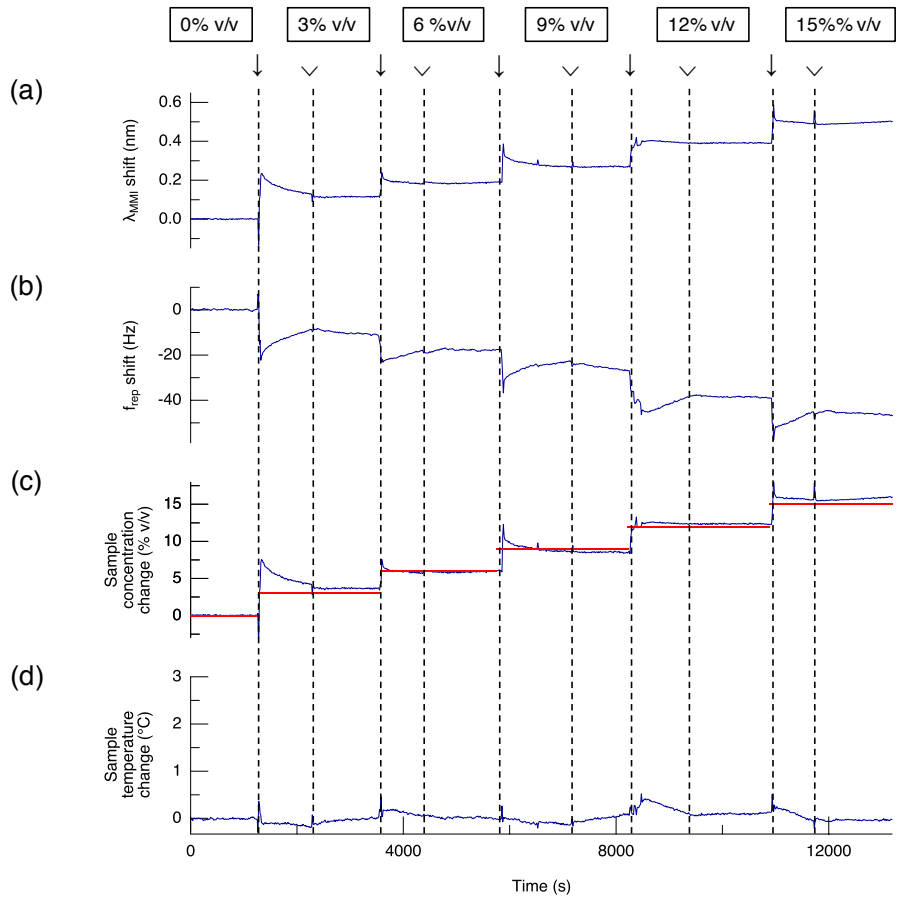

Fig. 5. Temporal behaviors of the $\lambda_{MMI}$ shift, $f_{rep}$ shift, the estimated sample concentration change and the estimated temperature change under the constant temperature (26.0ºC) and the variable ethanol concentration (0-15% v/v). The arrows indicate the times at which the ethanol was applied. The arrow heads indicate the times at which the sample was stirred.

We finally changed both the concentration and temperature of a sample. The protocol of the experiment is as follows:

[1] The ethanol concentration and temperature were initially set to 0% v/v and 27.0ºC, respectively.

[2] The ethanol temperature and concentration were simultaneously increased by 0.5ºC and 2.5% v/v, respectively.

[3] The ethanol temperature and concentration were simultaneously increased by 0.5ºC and 2.5% v/v, respectively.

[4] The ethanol temperature and concentration were simultaneously increased by 1.0ºC and 2.5% v/v, respectively.

[5] The ethanol concentration and temperature were simultaneously increased by 2.5% v/v and decreased by 0.5ºC, respectively.

These changes were sensitively reflected in $\Delta\lambda_{MMI}$ and $\Delta f_{rep}$, as shown in Figs. 6a and 6b, respectively. By using Eq. 3, $\Delta C$ and $\Delta T$ were determined, as shown in Figs. 6c and 6d. Both $\Delta C$ and $\Delta T$ could be determined with reasonable precisions even with the simultaneous changes in sample temperature and concentration. These results indicated the feasibility of the MMI-OFC sensing cavity for the temperature-compensated RI sensing or the simultaneous measurement of the material-depended RI and the sample temperature.

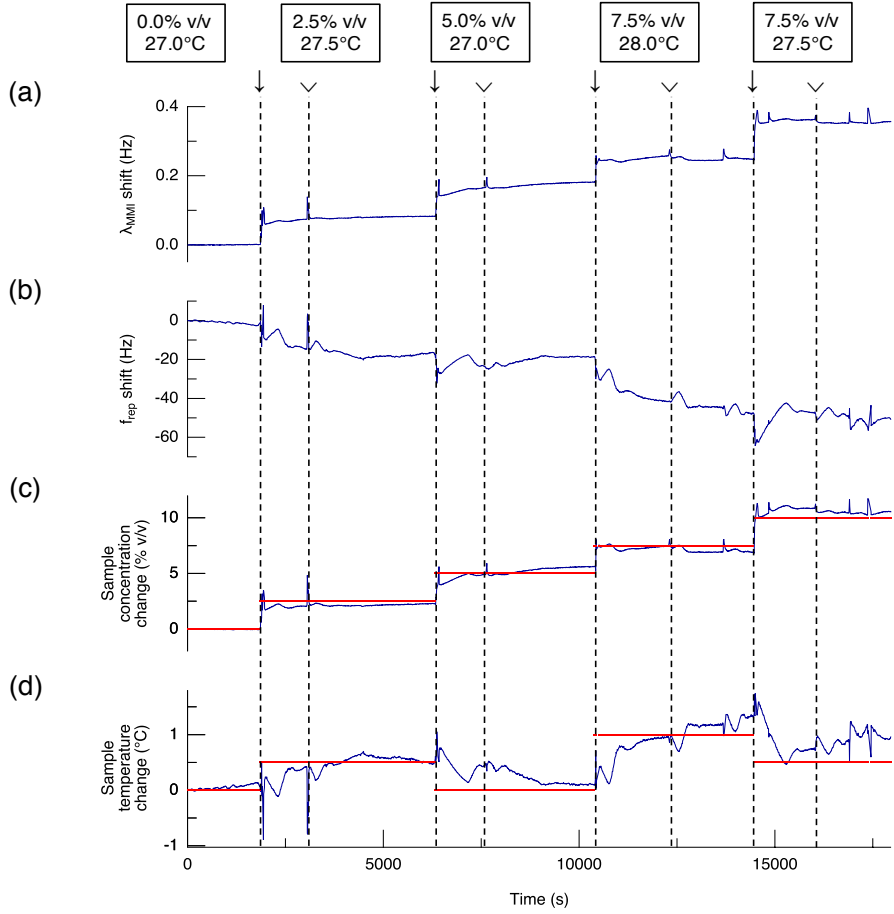

Fig. 6. Temporal behaviors of $\lambda_{MMI}$ shift, $f_{rep}$ shift, the estimated sample concentration change and the estimated temperature change under the variable temperature and the variable ethanol concentration. The arrows indicate the times at which the ethanol was applied. The arrow heads indicate the times at which the sample was stirred.

## 4. Discussion

In the proposed MMI-OFC sensing cavity, the RI sensing with temperature compensation was realized by observing the optical spectral shift $\Delta\lambda_{MMI}$ and the RF spectral shift $\Delta f_{rep}$. Our finding is that the material-dependent RI and the sample temperature can change both the $\Delta\lambda_{MMI}$ and $\Delta f_{rep}$, although with different slope coefficients. First, we discuss the detection mechanism to corroborate the regularity of the slope coefficient matrix in Eq. (2).

The interference wavelength $\lambda_{MMI}$ is determined as the following equation [9],

$$\lambda_{MMI} = \frac{n_{MMF} m}{L_{MMF}} D_{MMF} \left(n_{sam}\right)^2, \tag{3}$$

where $n_{MMF}$ is the RI of the MMF, $L_{MMF}$ is the geometrical length of the MMF, $D_{MMF}$ is the effective core diameter of the MMF, $n_{sam}$ is the RI of the target sample, and $m$ is the interference index of the MMI fiber sensor. The concentration of the sample can change the RI of the sample. The temperature of the sample can change not only the RI of the sample via the thermo-optic effect but also the RI of the MMF, the length of the MMF and the diameter of the MMF via the thermo-optic effect or the thermal expansion effect. Thus, the partial derivatives of $\lambda_{MMI}$ against concentration and temperature are given by,

$$\frac{\partial \lambda_{MMI}}{\partial C} = \frac{\partial \lambda_{MMI}}{\partial n_{sam}} \frac{\partial n_{sam}}{\partial C}, \tag{4}$$

$$\frac{\partial \lambda_{MMI}}{\partial T} = \frac{\partial \lambda_{MMI}}{\partial n_{sam}} \frac{\partial n_{sam}}{\partial T} + \frac{\partial \lambda_{MMI}}{\partial n_{MMF}} \frac{\partial n_{MMF}}{\partial T} + \frac{\partial \lambda_{MMI}}{\partial L_{MMF}} \frac{\partial L_{MMF}}{\partial T} + \frac{\partial \lambda_{MMI}}{\partial D_{MMF}} \frac{\partial D_{MMF}}{\partial T}. \tag{5}$$

In contrast, as shown in Eq. 1, the repetition frequency $f_{rep}$ is a function of cavity parameters, including the geometrical length of the fiber cavity, which would be modulated by the diameter and the length of the MMF, and the RI of the fiber cavity. The repetition frequency $f_{rep}$ is also a function of the RI of the sample because the RI-dependent wavelength shift of the MMI fiber sensor must change the optical length of the cavity via the wavelength dispersion of the cavity fiber. In the same manner as the interference wavelength $\lambda_{MMI}$, the partial derivatives of $f_{rep}$ against concentration and temperature are given by,

$$\frac{\partial f_{rep}}{\partial C} = \frac{\partial f_{rep}}{\partial n_{sam}} \frac{\partial n_{sam}}{\partial C}, \tag{6}$$

$$\frac{\partial f_{rep}}{\partial T} = \frac{\partial f_{rep}}{\partial n_{sam}} \frac{\partial n_{sam}}{\partial T} + \frac{\partial f_{rep}}{\partial n_{MMF}} \frac{\partial n_{MMF}}{\partial T} + \frac{\partial f_{rep}}{\partial L_{MMF}} \frac{\partial L_{MMF}}{\partial T} + \frac{\partial f_{rep}}{\partial D_{MMF}} \frac{\partial D_{MMF}}{\partial T}. \tag{7}$$

Therefore, the discriminant equation of the coefficient matrix of Eq. 2 can be expressed with Eqs. 4-7 as follows,

$$\begin{aligned}
D &= \frac{\partial f_{rep}}{\partial C}\frac{\partial \lambda_{MMI}}{\partial T} - \frac{\partial f_{rep}}{\partial T}\frac{\partial \lambda_{MMI}}{\partial C} \\
&= \frac{\partial n_{sam}}{\partial C}\left[\left(\frac{\partial f_{rep}}{\partial n_{sam}}\frac{\partial \lambda_{MMI}}{\partial n_{MMF}} - \frac{\partial f_{rep}}{\partial n_{MMF}}\frac{\partial \lambda_{MMI}}{\partial n_{sam}}\right)\frac{\partial n_{MMF}}{\partial T} + \left(\frac{\partial f_{rep}}{\partial n_{sam}}\frac{\partial \lambda_{MMI}}{\partial L_{MMF}} - \frac{\partial f_{rep}}{\partial L_{MMF}}\frac{\partial \lambda_{MMI}}{\partial n_{sam}}\right)\frac{\partial L_{MMF}}{\partial T}\right. \\
&\quad \left.\left(\frac{\partial f_{rep}}{\partial n_{sam}}\frac{\partial \lambda_{MMI}}{\partial D_{MMF}} - \frac{\partial f_{rep}}{\partial D_{MMF}}\frac{\partial \lambda_{MMI}}{\partial n_{sam}}\right)\frac{\partial D_{MMF}}{\partial T}\right].
\end{aligned} \quad (8)$$

According to the discriminant equation of the coefficient matrix, the importance of the detection mechanism is the presence of the temperature dependent changes of the MMI fiber sensor, *i.e.*, the RI, the geometrical length and the diameter of the MMF. Since our results indicated that the discriminant equation is not 0, in other words, the slope coefficient matrix is regular, the presence of the temperature dependent MMF change effectively utilized for the simultaneous measurement of the RI sensing with temperature compensation with the MMI-OFC sensing cavity.

We also discuss the validity of the slope coefficients of the $\Delta\lambda_{MMI}$ and the $\Delta f_{rep}$. These slope coefficients are determined by the concentration-dependent and temperature-dependent change of sample RI $n_{sam}$. In general, it is known that the water/ethanol mixture shows a concentration-dependent RI change on the order of $5\times10^{-4}$ (RIU/% v/v) [30] and temperature-dependent RI change on the order of $1\times10^{-4}$ (RIU/°C) [31]. Therefore, these RI changes are comparable to each other, as shown in our results. In our previous research, the RI sensitivity of the MMI sensor used in this experiment was 130.3 nm/RIU [13]. From these values, the concentration-dependent and temperature-dependent $\Delta\lambda_{MMI}$ and $\Delta f_{rep}$ slopes are, respectively, estimated as 0.061 nm/% v/v and 0.013 nm/°C, which are in reasonable agreement with the experimental data. An RI sensitivity of $-6.19\times10^{3}$ Hz/RIU was determined by RI-dependent $\Delta\lambda_{MMI}$ and the wavelength dispersion of the cavity fiber. Therefore, a concentration-dependent $\Delta f_{rep}$ with slope coefficient of -2.90 Hz/% v/v was estimated, which is comparable to the experimental value (= -2.96 Hz/% v/v). In the same manner, a concentration-dependent $\Delta f_{rep}$ with slope coefficient of -10.72 Hz/°C is calculated;

therefore, this value is reasonable in comparison with the experimental value (= -20.46 Hz/°C). This value is in reasonable agreement with measured values. The difference between them is mainly due to the temperature fluctuation of the optical fiber near the sensing region.

## 5. Conclusion

In conclusion, we proposed an RI sensing method with temperature compensation by using an MMI-OFC sensing cavity. We provided a proof-of-principle demonstration of the proposed method by the simultaneous measurement of temperature- and concentration-dependent spectral shifts of $\Delta\lambda_{MMI}$ and $\Delta f_{rep}$ slopes, realizing the simultaneous measurement of temperature and concentration of a liquid sample. We expected that the RI sensing with temperature compensation based on the MMI-OFC sensing cavity will be an effective way to precisely characterize and identify materials in various fields and opens a new aspect of OFC-based instrumentations.

## Funding

Exploratory Research for Advanced Technology (ERATO) MINOSHIMA Intelligent Optical Synthesizer Project (JPMJER1304), Japan Science and Technology Agency (JST), Japan; Grant-in-Aid for Exploratory Research (15K13384) from the Japan Society for the Promotion of Science (JSPS).

## Acknowledgments

The authors thank Ms. Natsuko Takeichi and Ms. Shoko Lewis, Tokushima University, for the English proofreading of the manuscript.

# References

[1] R. Budwig, "Refractive index matching methods for liquid flow investigations," Exp. Fluids **17**(5), 350-355 (1994).

[2] C. Wu, A. B. Khanikaev, R. Adato, N. Arju, A. A. Tanik, H. Altug, and G. Shvets, "Fano-resonant asymmetric metamaterials for ultrasensitive spectroscopy and identification of molecular monolayers," Nat. Mater. **11**(1), 69-75 (2012).

[3] A. V. Kabashin, P. Evans, S. Pastkovsky, W. Hendren, G. A. Wurtz, R. Atkinson, R. Pollard, V. A. Podolskiy and A. V. Zayats, "Plasmonic nanorod metamaterials for biosensing," Nat. Mater. **8**(11), 867-871 (2009).

[4] S. K. Chauhan, N. Punjabi, D. K. Sharma, and S. Mukherji, "A silicon nitride coated LSPR based fiber-optic probe for possible continuous monitoring of sucrose content in fruit juices," Sens. Actuators B Chem. **222**, 1240-1250 (2016).

[5] B. Sutapun, M. Tabib-Azar, and A. Kazemi, "Pd-coated elastooptic fiber optic Bragg grating sensors for multiplexed hydrogen sensing," Sens. Actuators B Chem. **60**(1), 27-34 (1999).

[6] J. R. Ott, M. Heuck, C. Agger, P. D. Rasmussen, and O. Bang, "Label-free and selective nonlinear fiber-optical biosensing," Opt. Express **16**(25), 20834-20847 (2008).

[7] D. Monzón-Hernández, J. lVillatoro, "High-resolution refractive index sensing by means of a multiple-peak surface plasmon resonance optical fiber sensor," Sens. Actuators B Chem. **115**(1), 227-231 (2006).

[8] G, Yin, S. Lou, and H. Zou, "Refractive index sensor with asymmetrical fiber Mach–Zehnder interferometer based on concatenating single-mode abrupt taper and core-offset section," Opt. Laser Technol. **45**, 294-300 (2013).

[9] H. Fukano, T, Aiga, and S. Taue, "High-sensitivity fiber-optic refractive index sensor based on multimode interference using small-core single-mode fiber for biosensing," Jpn. J. Appl. Phys. **53**(4S), 04EL08 (2014).

[10] Y. Yang, M. Wang, Y. Shen, Y. Tang, J. Zhang, Y. Wu, S. Xiao, J. Liu, B. Wei, Q. Ding, and S. Jian, "Refractive index and temperature sensing based on an optoelectronic oscillator incorporating a Fabry-Perot fiber Bragg grating," IEEE Photon. J. **10**(1), 6800309 (2018).

[11] J. Zhang, Q. Sun, R. Liang, J. Wo, D. Liu, and P. Shum, "Microfiber Fabry–Perot interferometer fabricated by taper-drawing technique and its application as a radio frequency interrogated refractive index sensor," Opt. Lett. **37**(14), 2925-2927 (2012).

[12] L. Stern, A. Naiman, G. Keinan, N. Mazurski, M. Grajower, and U. Levy, "Ultra-precise optical to radio frequency based chip-scale refractive index and temperature sensor," Optica **4**(1), 1-7 (2017).

[13] R. Oe, S. Taue, T. Minamikawa, K. Nagai, K. Shibuya, T. Mizuno, M. Yamagiwa, Y. Mizutani, H. Yamamoto, T. Iwata, H. Fukano, Y. Nakajima, K. Minoshima, and T. Yasui, "Refractive-index-sensing optical comb based on photonic radio-frequency conversion with intracavity multi-mode interference fiber sensor," Opt. Express **26**(15), 19694-19706 (2018).

[14] Th. Udem, J. Reichert, R. Holzwarth, and T. W. Hänsch, "Accurate measurement of large optical frequency differences with a mode-locked laser," Opt. Lett. **24**(13), 881–883 (1999).

[15] Th. Udem, R. Holzwarth, and T. W. Hänsch, "Optical frequency metrology," Nature **416**, 233–237 (2002).

[16] M. Niering, R. Holzwarth, J. Reichert, P. Pokasov, Th. Udem, M. Weitz, T. W. Hänsch, P. Lemonde, G. Santarelli, M. Abgrall, P. Laurent, C. Salomon, and A. Clairon, "Measurement

of the hydrogen 1S-2S transition frequency by phase coherent comparison with a microwave cesium fountain clock," Phys. Rev. Lett. **84**(24), 5496–5499 (2000).

[17] S. A. Diddams, L. Hollberg, and V. Mbele, "Molecular fingerprinting with the resolved modes of a femtosecond laser frequency comb," Nature **445**(6877), 627–630 (2007).

[18] A. Asahara, A. Nishiyama, S. Yoshida, K. Kondo, Y. Nakajima, and K. Minoshima, "Dual-comb spectroscopy for rapid characterization of complex optical properties of solids," Opt. Lett. **41**(21), 4971–4974 (2016).

[19] T. Minamikawa, Y.-D. Hsieh, K. Shibuya, E. Hase, Y. Kaneoka, S. Okubo, H. Inaba, Y. Mizutani, H. Yamamoto, T. Iwata, and T. Yasui, "Dual-comb spectroscopic ellipsometry," Nature Commun. **8**(1), 610 (2017).

[20] N. Kuse, A. Ozawa, and Y. Kobayashi, "Static FBG strain sensor with high resolution and large dynamic range by dual-comb spectroscopy," Opt. Express **21**(9), 11141–11149 (2013).

[21] T. Minamikawa, T. Ogura, Y. Nakajima, E. Hase, Y. Mizutani, H. Yamamoto, K. Minoshima, and T. Yasui, "Strain sensing based on strain to radio-frequency conversion of optical frequency comb," Opt. Express **26**(8), 9484-9491 (2018).

[22] T. Minamikawa, T. Masuoka, T. Ogura, K. Shibuya, R. Oe, E. Hase, Y. Nakajima, Y. Yamaoka, T. Mizuno, M. Yamagiwa, Y. Mizutani, H. Yamamoto, T. Iwata, K. Minoshima, and T. Yasui, "Ultrasonic wave sensing using an optical-frequency-comb sensing cavity for photoacoustic imaging," OSA Continuum **2**(2), 439-449 (2019).

[23] K. Minoshima and H. Matsumoto, "High-accuracy measurement of 240-m distance in an optical tunnel by use of a compact femtosecond laser," Appl. Opt. **39**(30), 5512-5517 (2000).

[24] I. Coddington, W. C. Swann, L. Nenadovic, and N. R. Newbury, "Rapid and precise absolute distance measurements at long range," Nature Photon. **3**(6), 351–356 (2009).

[25] T. Ideguchi, S. Holzner, B. Bernhardt, G. Guelachvili, N. Picqué, and T. W. Hänsch, "Coherent Raman spectro-imaging with laser frequency combs," Nature **502**, 355-358 (2013).

[26] E. Hase, T. Minamikawa, T. Mizuno, S. Miyamoto, R. Ichikawa, Y.-D. Hsieh, K. Shibuya, K. Sato, Y. Nakajima, A. Asahara, K. Minoshima, Y. Mizutani, T. Iwata, H. Yamamoto, and T. Yasui, "Scan-less confocal phase imaging based on dual-comb microscopy," Optica **5**(5), 634-643 (2018).

[27] E. Hase, T. Minamikawa, S. Miyamoto, R. Ichikawa, Y.-D. Hsieh, Y. Mizutani, T. Iwata, H. Yamamoto, and T. Yasui, "Scan-Less, Kilo-Pixel, Line-Field Confocal Phase Imaging with Spectrally Encoded Dual-Comb Microscopy," IEEE J. Sel. Top. Quant. **25**(1), 2879017 (2019).

[28] K. Shibuya, T. Minamikawa, Y. Mizutani, H. Yamamoto, K. Minoshima, T. Yasui, and T. Iwata, "Scan-less hyperspectral dual-comb single-pixel-imaging in both amplitude and phase," Opt. Express **25**(18), 21947-21957 (2017).

[29] T. Kato, M. Uchida, and K. Minoshima, "No-scanning 3D measurement method using ultrafast dimensional conversion with a chirped optical frequency comb," Sci. Rep. **7**, 3670 (2017).

[30] J. V. Herráez, and R. Belda, "Refractive Indices, Densities and Excess Molar Volumes of Monoalcohols + Water," J. Solution Chem. **35**(9), 1315-1328 (2006).

[31] A. N. Bashkatov, and E. A. Genina, "Water refractive index in dependence on temperature and wavelength: a simple approximation," Proc. SPIE **5068**, 393-396 (2003).